\documentclass[10pt,twocolumn,letterpaper]{article}

\usepackage[pagenumbers]{iccv} 

\definecolor{iccvblue}{rgb}{0.21,0.49,0.74}
\usepackage[pagebackref,breaklinks,colorlinks,allcolors=iccvblue]{hyperref}
\definecolor{gray}{rgb}{0.92, 0.92, 0.92}
\definecolor{yellow}{rgb}{1, 1, 0.7}
\definecolor{orange}{rgb}{1, 0.85, 0.7}
\definecolor{red}{rgb}{1, 0., 0.}

\usepackage{multirow}
\usepackage{graphicx}
\usepackage{booktabs}
\usepackage{enumitem}
\newcommand\B[1]{\textcolor{blue}{#1}}
\AtBeginDocument{%
  }
\def\paperID{*****} 
\def\confName{ICCV}
\def\confYear{2025}

\title{iDiT-HOI: Inpainting-based Hand Object Interaction Reenactment via Video Diffusion Transformer}

\author{
Zhelun Shen$^1$, Chenming Wu\textsuperscript{\rm 1}\thanks{Corresponding author.}, Junsheng Zhou$^{1,2}$, Chen Zhao$^1$, Kaisiyuan Wang$^1$, \\Hang Zhou$^1$, Yingying Li$^1$, Haocheng Feng$^1$, Wei He$^1$, Jingdong Wang$^1$ \\ \vspace{-12pt} \\ 
\small $^1$Department of Computer Vision Technology(VIS), Baidu Inc., China \\
\small $^2$School of Software, Tsinghua University, China\\
\tt \small
\{shenzhelun, wuchenming\}@baidu.com
}

\begin{document}
\twocolumn[{%
\renewcommand\twocolumn[1][]{#1}%
\maketitle
}]
\begin{abstract}
Digital human video generation is gaining traction in fields like education and e-commerce, driven by advancements in head-body animation and lip-syncing technologies. However, realistic Hand-Object Interaction (HOI)—the complex dynamics between human hands and objects—continues to pose challenges. Generating natural and believable HOI reenactments is difficult due to issues such as occlusion between hands and objects, variations in object shapes and orientations, and the necessity for precise physical interactions, and importantly, the ability to generalize to unseen humans and objects.
This paper presents a novel framework \emph{iDiT-HOI} that enables in-the-wild HOI reenactment generation. Specifically, we propose a unified inpainting-based token process method, called \emph{Inp-TPU}, with a two-stage video diffusion transformer (DiT) model. The first stage generates a key frame by inserting the designated object into the hand region, providing a reference for subsequent frames. The second stage ensures temporal coherence and fluidity in hand-object interactions. The key contribution of our method is to reuse the pretrained model's context perception capabilities without introducing additional parameters, enabling strong generalization to unseen objects and scenarios, and our proposed paradigm naturally supports long video generation.
Comprehensive evaluations demonstrate that our approach outperforms existing methods, particularly in challenging real-world scenes, offering enhanced realism and more seamless hand-object interactions.


\end{abstract}    
\section{Introduction}

Digital human broadcasting videos play a pivotal role in contemporary society, particularly in light of the increasing demand for digital human content across various domains, including education and e-Commerce. Recent advancements in digital human technology have significantly enhanced capabilities in head-body animation and lip-syncing~\cite{wav2lip,lsp,vprq,liveportrait,pcavs,LIA,facevid2vid,vpgc,emo,hallo}, thereby improving the realism and expressiveness of virtual characters.
Nonetheless, a challenging area remains Hand-Object Interaction (HOI), which encompasses the intricate dynamics between human hands and objects. Effective HOI is essential for fostering believable digital humans, as it contributes substantially to the overall authenticity and immersive experience of animated characters.

The generation of realistic hand movements presents several challenges. First, the diversity of object shapes and orientations complicates the task, as the hand's interaction with various objects requires a nuanced understanding of their physical properties. Additionally, achieving natural and plausible hand motions is crucial for maintaining viewer engagement and suspension of disbelief. Traditional animation techniques often falter in capturing these complexities, leading to disjointed interactions that can detract from the overall realism of the digital representation. As a result, there is a pressing need for innovative methodologies that can effectively model and generate realistic HOI.

In recent years, object-centric and human-centric Hand-Object Interaction (HOI) reenactment has garnered significant attention and progress. Notably, HOI-Swap~\cite{hoi-swap} extends image-level HOI inpainting to a video framework by incorporating sequential frame warping. However, it is limited to generating object-centric videos with single-hand grasping actions and exhibits poor generalization to novel objects. More recently, Re-HOLD~\cite{fan2025re} introduced a layout-instructed diffusion model for HOI reenactment. While this layout video aids training convergence, it requires additional human input during inference to curate control videos for natural hand-object interactions, relying on rendered hand meshes and layout videos. Moreover, it duplicates the main diffusion network for object identification preservation, resulting in increased parameters. 

This work addresses the challenges in high-fidelity HOI reenactment video generation from a source video and a reference object. Our objective is to enhance HOI reenactment beyond source domain datasets and extend it to challenging in-the-wild scenarios. We propose a paradigm that facilitates HOI reenactment while reusing established context perception capabilities of existing video generation models without introducing new parameters. To achieve this, we introduce a unified token processing method that leverages attention parameters in masked regions, a feature often overlooked in previous approaches that rely on additional attention modules or parameter-copied reference networks. This significantly improves the model's generalization to in-the-wild tasks, distinguishing our method from others.
Furthermore, we note that existing approaches, such as ReHOLD, rely heavily on external control signals to transfer original motion for adapting to new objects. This reliance on human input can result in unnatural hand-object interactions.
Our solution is to design a model akin to video inpainting, enhanced by a strong identity constraint and our unified token processing method, allowing the network to generate more natural and plausible HOI clips in the target masked regions, thereby improving the overall quality of the reenactment video.

Specifically, we design a two-stage DiT model for HOI reenactment video generation. The first stage focuses on precisely inserting the designated object into the hand region, producing a key frame that establishes the initial interaction. This key frame serves as a crucial reference for subsequent frame generation, ensuring contextual relevance of hand movement. The second stage then generates the remaining frames, maintaining coherence and fluidity in the hand-object interaction. Both stages are trained using an inpainting-based methodology, which allows for a more integrated learning process that accounts for the complexities of HOI.
We conduct comprehensive qualitative and quantitative evaluations to validate our approach, comparing it against existing methods in the field. Our results demonstrate that the proposed framework significantly outperforms traditional techniques, providing superior realism and fluidity in hand movements and interactions. By addressing the intricacies of HOI, our work not only advances the state of digital human animation but also highlights the importance of incorporating realistic hand-object interactions to create immersive virtual experiences.
We summarize the contributions of this paper as follows.

\begin{itemize}
    \item We introduce a unified inpainting-based token processing method within the DiT video generation architecture, which effectively accounts for hand movements and object details, ensuring natural contact and interactions between them.
    \item Our method efficiently reuses the established context perception capabilities of the pretrained model without adding new trainable parameters, which improves performance in real-world HOI generalization tasks, such as object swapping, and facilitates seamless exchanges.
    \item Extensive experiments on self-reenactment and cross-reenactment—especially in challenging in-the-wild scenarios—demonstrate the effectiveness of our proposed method, highlighting its potential for broad application in industrial-scale HOI video generation.
\end{itemize}


\section{Related Work}

\subsection{Human Body Animation}
The rapid advancement of diffusion models—especially the impact of open-source models like Stable Diffusion—has catalyzed significant progress in addressing the challenges of human body animation. A prominent line of work leverages U-Net architectures~\cite{unet} enhanced with various cross-attention mechanisms~\cite{attention} to incorporate additional signals, such as contextual or textural information, into the animation generation process. One of the pioneering efforts in this space is PIDM~\cite{pidm}, which introduces classifier-free diffusion guidance to synthesize human images from pose inputs. Building on this foundation, DreamPose~\cite{dreampose} utilizes UV maps as motion signals and applies conditional embeddings for motion transfer. This has inspired a wave of concurrent works—including AnimateAnyone~\cite{animateanyone}, DISCO~\cite{disco}, MimicMotion~\cite{zhang2024mimicmotion}, MagicPose~\cite{magicpose}, ShowMaker~\cite{yang2025showmaker}, and TalkAct~\cite{talkact}—that explore similar ideas.
More recently, RealisDance~\cite{realisdance} explores the integration of multiple conditioning inputs, such as DW-Pose~\cite{dwpose}, SMPL~\cite{smplx}, and HaMeR~\cite{hamer}, to better capture nuanced hand movements by exploiting the strengths of diverse human body representations. Some innovative approaches have begun extending traditional condition-driven animation to composable human animation, where background scenes and foreground objects are incorporated for tasks like personalized human insertion~\cite{men2024mimo, hu2025animate}.
Following the scalability demonstrated by diffusion transformers (DiT), recent efforts have adopted DiT-based architectures for human animation. Notable examples include HumanDiT~\cite{gan2025humandit}, Human4DiT~\cite{shao2024human4dit}, UniAnimate~\cite{wang2025unianimate}, and DreamActor-M1~\cite{luo2025dreamactor}.

\subsection{Diffusion-based Video Editing and Inpainting}
A substantial body of research has focused on adapting pre-trained diffusion models for video generation, with extensions to editing and inpainting tasks. These efforts include both zero-shot methods—requiring no additional training for new inputs—such as~\cite{tokenflow, rerender, pix2video, fatezero, space, vidtome, wang2023diffusion}, and one-shot-tuned frameworks like~\cite{tuneavideo, cut, shape, videoswap}, which require lightweight fine-tuning on individual video clips. While these techniques achieve promising results, they often rely on per-video adaptation, introducing additional training overhead beyond the inference stage, which limits their scalability and practical deployment.
To address this, alternative approaches have emerged that employ training-based strategies designed for generalization. These methods—including Structure-A-Video~\cite{structure}, Control-A-Video~\cite{control}, FlowVid~\cite{flowvid}, VideoComposer~\cite{videocomposer}, MoonShot~\cite{moonshot}, VASE~\cite{vase}, OmniMatte~\cite{omnimatte}, Zhu et al.\cite{zhu2022discrete}, and~\cite{liu2024iterative}—train models on large-scale video datasets, enabling fast and versatile inference-time editing without the need for per-video tuning. However, these models struggle with fine-grained generation, especially in complex human-object interactions (HOIs), which remain a key challenge in video synthesis.

Many video generation and editing frameworks~\cite{stablediffusion, tokenflow, emu, rerender, fatezero, pix2video, space, tuneavideo, shape, ccedit, editavideo, videop2p, makeavideo, videogeneration} rely heavily on textual prompts for control. While convenient, text-based guidance often lacks the precision needed to describe object details or spatial relationships, especially in complex human-object interaction scenarios.
Recent diffusion-based video inpainting methods such as AVID~\cite{zhang2024avid} and COCOCO~\cite{zi2025cococo} build on Stable Diffusion Inpainting~\cite{sdinp} by adding trainable temporal attention layers. VideoPainter~\cite{bian2025videopainter} further advances the field with a DiT-based dual-branch architecture that separates foreground generation from background preservation.
Despite their strengths, these methods rely on text inputs, which can be ambiguous. Our approach uses object images for more precise control, improving realism and alignment with user intent in human-object interaction generation.
Moreover, a concurrent work to ours is VACE~\cite{jiang2025vace}, a general-purpose, all-in-one model for video creation and editing that supports various condition combinations—including image-reference and inpainting—similar to our approaches, but relies on a dual-stream architecture with duplicated parameters. In contrast, our method reuses masked-region parameters for more efficient training and better generalization, outperforming VACE in HOI tasks.


\begin{figure*}[t]
\includegraphics[width=\linewidth]{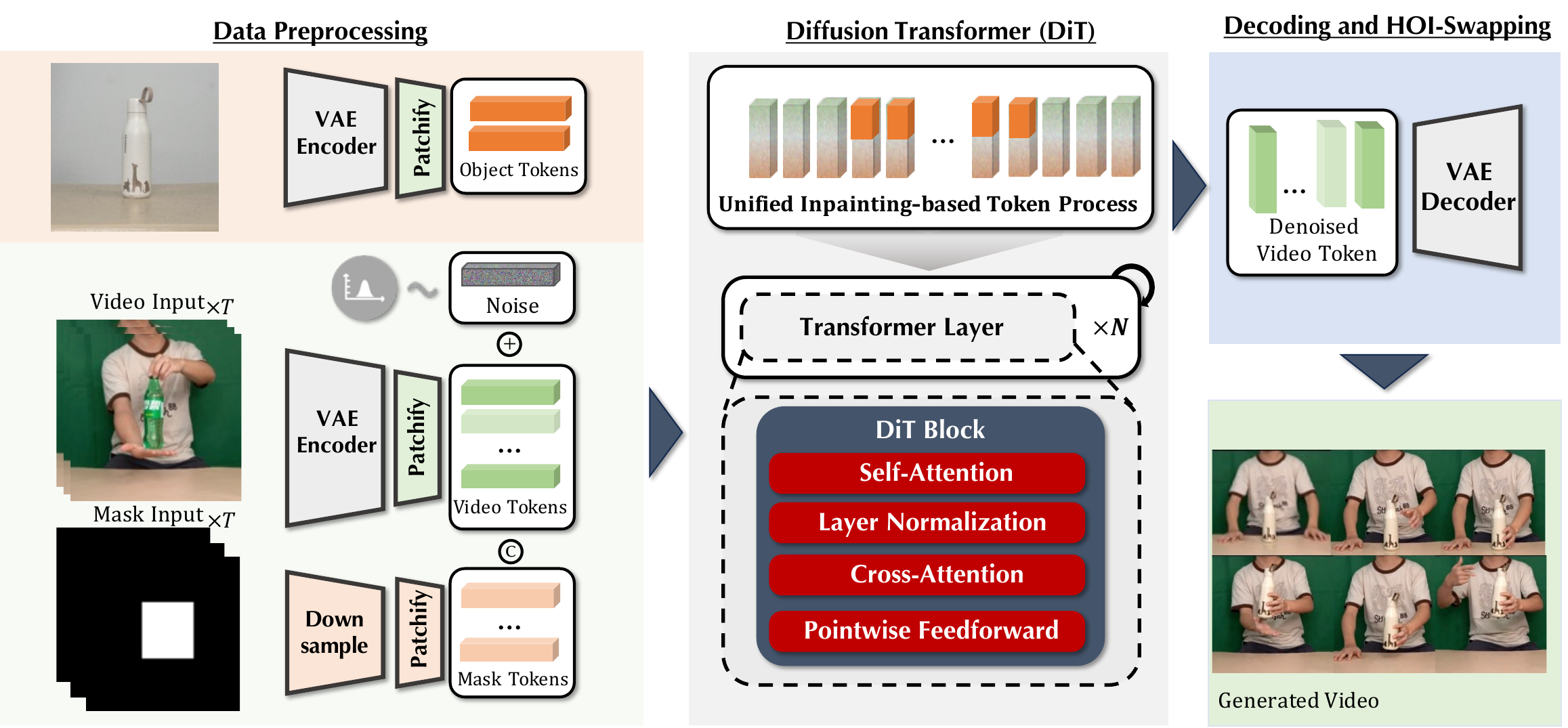}
\caption{
Our proposed method begins with Data Preprocessing, where object images and input video frames with corresponding masks are encoded into latent tokens and patchified. These tokens are then fed into the Diffusion Transformer (DiT), which employs a unified inpainting-based token process and iterative transformer layers to denoise and refine the video tokens. Finally, the denoised video tokens are decoded back into video frames using the VAE decoder, producing the final generated videos. The whole DiT model is trained in a self-supervised (reconstruction) manner, and the reference objects can be easily replaced during inference.}
\label{fig:pipeline}
\end{figure*}

\subsection{Human-Object Interaction}
Human motion is inherently contextual, often occurring in conjunction with objects and within structured environments. To address this, various methods have been proposed to realistically integrate humans into scenes~\cite{place, synthesizing, populating, generating, resolving}. In parallel, significant attention has been given to modeling fine-grained hand-object interactions (HOIs)~\cite{hoi, affordance, diffusionhoi, hoidiffusion, cghoi, graspxl, lego}, which are critical for achieving realistic human behavior in both static and dynamic settings.
Among these, DiffHOI~\cite{diffusionhoi} employs a diffusion-based framework to model the conditional distribution of object renderings, thereby facilitating novel-view synthesis of HOIs. GraspXL~\cite{graspxl} offers a unified framework for generating hand-object grasping motions across a range of object geometries and hand morphologies. Cg-HOI~\cite{cghoi} focuses on generating plausible 3D human-object interactions from textual descriptions, enhancing controllability in semantic-to-motion generation. HOI-Swap~\cite{hoi-swap} extends static image-level HOI inpainting to a video-based framework by incorporating sequential frame warping. However, it is constrained to object-centric scenes involving single-hand grasping, and its ability to generalize to unseen objects remains limited.
More recently, Re-HOLD~\cite{fan2025re} introduced a layout-instructed diffusion model for HOI reenactment, using layout videos as auxiliary inputs to accelerate training convergence. Despite this, Re-HOLD requires additional user input at inference time, including curated layout videos and rendered hand meshes, to ensure realistic hand-object interactions. Furthermore, it duplicates the main diffusion network to preserve object identity, resulting in a significant increase in model parameters and computational overhead.

\section{Proposed Method}

\subsection{Preliminary}
\label{sec:method:preliminary}

\noindent \textbf{Diffusion Transformer (DiT)} models are utilized in various generation tasks such as image generation architectures FLUX.1 \cite{flux2024}, Stable Diffusion 3~\cite{esser2024scaling}, PixArt \cite{chen2023pixart}, and HunyuanDiT~\cite{li2024hunyuan}, as well as in video generation architectures of SORA \cite{sora}, HunyuanVideo~\cite{kong2024hunyuanvideo}, and Wan 2.1~\cite{wan2025wanopenadvancedlargescale}. These models employ transformers as denoising networks to iteratively refine noisy tokens. A DiT model processes two types of tokens: noisy tokens \( \mathbf{X} \in \mathbb{R}^{N \times d} \) and text condition tokens \( \mathbf{C}_T \in \mathbb{R}^{M \times d} \), where \( d \) is the embedding dimension, and \( N \) and \( M \) represent the number of image/video and text tokens, respectively. Throughout the network, these tokens maintain consistent shapes as they pass through multiple transformer blocks, as shown in the mid part of Fig.~\ref{fig:pipeline}.

\subsection{Overview}
\label{sec:method:overview}

The goal of hand-object interaction (HOI) reenactment is to synthesize realistic and coherent interactions between hands and a target object, given a masked input video and an object reference image \( I_{\text{ref}} \in \mathbb{R}^{H \times W \times 3} \). Our framework is trained in a self-supervised reconstruction manner, where both the masked video \( \tilde{\mathbf{V}} \in \mathbb{R}^{F \times H \times W \times 3} \) and the reference image \( I_{\text{ref}} \) are derived from the original source video \( \mathbf{V} = \{ I_1, I_2, \ldots, I_F \} \in \mathbb{R}^{F \times H \times W \times 3} \).
The training objective is to reconstruct \( \mathbf{V} \) from random noise conditioned on \( \tilde{\mathbf{V}} \) and \( I_{\text{ref}} \), formulated as:
\[
\mathbf{V} = \mathcal{D}(\mathcal{M}(\mathbf{X}_\text{rand}, \mathcal{E}(I_\text{ref}), \mathcal{E}(\tilde{\mathbf{V}}))),
\]
where \( \mathbf{X}_\text{rand} \) represents randomly initialized noise tokens. We leverage a pretrained VAE for video compression, with encoder \( \mathcal{E} \) and decoder \( \mathcal{D} \). The input video \( \mathbf{V} \) is compressed into latent features \( \mathbf{Z} = \mathcal{E}(\mathbf{V}) \), and reconstruction is performed in the latent space after patchifying $\mathbf{Z}$ to $\mathbf{X}$ using patch embedding.

The denoising module \( \mathcal{M} \) follows the DiT architecture (see Sec.~\ref{sec:method:preliminary}), with our key contribution being a unified inpainting-based token conditioning mechanism. This design effectively incorporates visual information from the reference image while reusing the pretrained DiT's context-aware capabilities. The output tokens are reshaped to match the latent dimensions and decoded to reconstruct the final video. Our approach adopts a two-stage generation strategy: we first employ a DiT-based model \( \mathcal{M}_\text{img} \) to generate a key frame \( I^\text{key} \) corresponding to the first frame of the output video. Subsequently, a separate DiT model \( \mathcal{M}_\text{vid} \) generates the remaining frames, conditioned on \( I^\text{key} \).
After full fine-tuning, the model generalizes well across multiple HOI tasks, including self-reenactment and cross-object reenactment. Moreover, it performs effectively in in-the-wild HOI swapping scenarios—where many prior methods require per-video fine-tuning and struggle to generalize. During inference, a novel object reference image \( I_\text{ref}' \) is supplied to reenact a target video \( \mathbf{V}' \). Layout modifications are applied to adapt the hand and object bounding boxes according to object differences, while hand poses are preserved to ensure realistic and adaptive interaction. An overview of the pipeline is shown in Fig.~\ref{fig:pipeline}.



\subsection{Unified Inpainting-based Token Process}
Unlike general video inpainting methods that utilize fine-grained masks to preserve background details, HOI reenactment often involves replacing one object with another of markedly different shape. In this context, object masks are typically represented using standardized methods like oriented bounding boxes (OBBs), requiring the model to enlarge the background while maintaining object identity. Conventional inpainting models focus on generating content within precisely defined masked regions, primarily ensuring temporal identity preservation. Most of the existing methods for reference injection, such as~\cite{jiang2025vace} additional parameters. In contrast, we propose a more efficient unified inpainting-based token process method that directly inserts the object into the masked region, reusing existing attention parameters on masked tokens for HOI reenactment video generation.

The core element of our HOI reenactment framework is the unified inpainting-based token processing unit -- \emph{Inp-TPU}. The design of Inp-TPU thoughtfully incorporates the insight of reusing the established context perception framework during the training of the generation model, without introducing additional trainable parameters. This approach enhances performance in real-world generalization tasks of HOI, such as object swapping, and enables seamless object exchanges. Our design philosophy focuses on utilizing the contextual perception capabilities of a pretrained I2V model to inject object information into the masked inpainting regions across the frames of a given frame or the whole video. 

Specifically,
given the input video $\tilde{\mathbf{V}}$, reference object $I_\text{ref}$, and mask $\mathbf{M}$ for compiling $\tilde{\mathbf{Z}}$, our Inp-TPU first extends $I_\text{ref}$ along its temporal axis to match the number of frames of $\tilde{\mathbf{V}}$, and obtains $\mathbf{V}_\text{ref}$. Then we use $\mathbf{M}$ to transform $\mathbf{V}_\text{ref}$ based on the centroid and sizes of the masked regions to obtain a video $\mathbf{V}^*_\text{ref}$ that not only temporally but also spatially aligned with $\tilde{\mathbf{V}}$. Those two videos are encoded by the VAE to obtain $\tilde{\mathbf{Z}}$ and $\mathbf{Z}^*_\text{ref}$, and being patchified to tokens $\tilde{\mathbf{X}}$ and $\mathbf{X}^*_\text{ref}$. Meanwhile, we downsamples $\mathbf{M}$ using bilinear interpolation to match the latent shape of $\tilde{\mathbf{Z}}$, and then applies a reshaping operation follows the intrinsic shape transform principle of the patch embedding to obtain the tokens of mask, denoted as $\mathbf{X}_\mathbf{M} \in [0, 1]$. Subsequently, we obtain the final conditional tokens $\mathbf{X}_\text{cond}$ to be concatenated with the noise tokens $\mathbf{X}_\text{rand}$ using the following simple yet effective formula:

\begin{equation}
    \mathbf{X}_\text{cond} = (1-\mathbf{X}_\mathbf{M})\cdot \tilde{\mathbf{X}} + \mathbf{X}_\mathbf{M}\cdot \mathbf{X}^*_\text{ref}.
\end{equation}
In this context, we use video for explanation purposes, as an image can be viewed as a special case of video consisting of a single frame. Therefore, our proposed inpainting-based token process can be applied at both video and image levels, as a unified process unit.




\subsection{HOI Reenactment Generation}

\noindent \textbf{Key Frame Generation.} Our method adheres to the image-then-video generation paradigm, similar to the approach outlined in~\cite{hoi-swap}. In the generation of key frame $I^\text{key}$, our DiT model executes in an image-to-image manner (I2I) and processes a shorter sequence of tokens through the proposed Inp-TPU module, which enhances efficiency and facilitates more effective handling of temporal information.
To improve the quality of the reference used in the generation process, we leverage the self-reconstruction strategy introduced in~\cite{mao2025ace++}. This approach allows for the simultaneous reconstruction of both the reference image and the key frame, ensuring that the generated content maintains high fidelity and coherence with the original visual context.

\vspace{4pt} \noindent \textbf{Subsequent Frame Generation.} Building upon the key frame $I^\text{key}$ generated by our DiT model $\mathcal{M}_\text{img}$ utilizing the Inp-TPU module, we subsequently integrate this frame within our DiT model $\mathcal{M}_\text{vid}$  in a video-to-video (V2V) context. This process not only maintains the use of the Inp-TPU module but also plays a critical role in the effective integration of reference tokens, thereby ensuring that the resulting video output is both cohesive and contextually relevant.
In a manner analogous to the key frame generation process described earlier, the latent representations of the source video and the corresponding mask are concatenated prior to being patchified into tokens. This concatenation facilitates the incorporation of essential information from both the source video and the mask. Additionally, the reference tokens are integrated into the denoising process of these tokens using the Inp-TPU module, further enhancing the quality of the generated video. This systematic approach not only optimizes the generation of sequential frames but also contributes to the overall coherence and fidelity of the output video.

\vspace{4pt} \noindent \textbf{Long Video Generation.} Considering the balance between training resources and efficiency requirements, a typical DiT video generation model processes approximately one hundred frames simultaneously, specifically denoising a token of temporally compressed latent data across these frames. Given that standard frame rates on the internet range from 25 to 30 frames per second, the resulting videos are often too brief for applications that necessitate longer and more continuous sequences.
Fortunately, our method facilitates subsequent frame generation in a V2V manner. This allows us to utilize the last frame of a preceding video clip as the initial image frame for generating a longer video sequence, without the need to train or fine-tune a dedicated long video generation model. Consequently, our approach enables the efficient production of extended video content for long video generation.



\subsection{Adaptive Masking for Shape Control}
Our approach necessitates a carefully crafted masking strategy for shape control. For instance, if we aim to swap a vertically oriented object with a horizontally oriented one.
Here, we propose an adaptive masking approach to introduce target object control into video generation.
Given a source image of the first video frame $I_1$ and its corresponding mask $\mathbf{M}_1$,  our goal is to generate an adaptive mask $\mathbb{M}_1$ that defines the placement region for the target object $I_{\text{ref}}$. The design of $\mathbf{M}_1$ adheres to the following three principles: (1) The generated mask $\mathbf{M}_1$ must fully encompass the original object in $I_1$. (2) The mask should be soft, as the inpainting models employed in our framework perform better with soft masks than with rigid binary bounding boxes. (3) The mask should maintain the same aspect ratio as the target object $I_{\text{ref}}$ to support scale-aware inpainting and prevent geometric distortion.

To satisfy these constraints, we design a multi-stage adaptive mask generation process. We first compute the oriented bounding box (OBB) of the mask $\mathbf{M}_1^{OBB}$ from the source mask $\mathbf{M}_1$, with height $h$ and width $w$. Next, we generate a soft ellipsoid mask whose semi-major and semi-minor axes are initially set to $h/2$ and $w/2$, respectively. To ensure that the ellipse fully covers the oriented bounding box, we enlarge both axes by a factor of $\sqrt{2}$. Finally, we adjust the aspect ratio of the ellipse to match that of the target object $I_{\text{ref}}$, enabling precise placement and shape adaptation.





\begin{figure*}[t]
\includegraphics[width=0.95\linewidth]{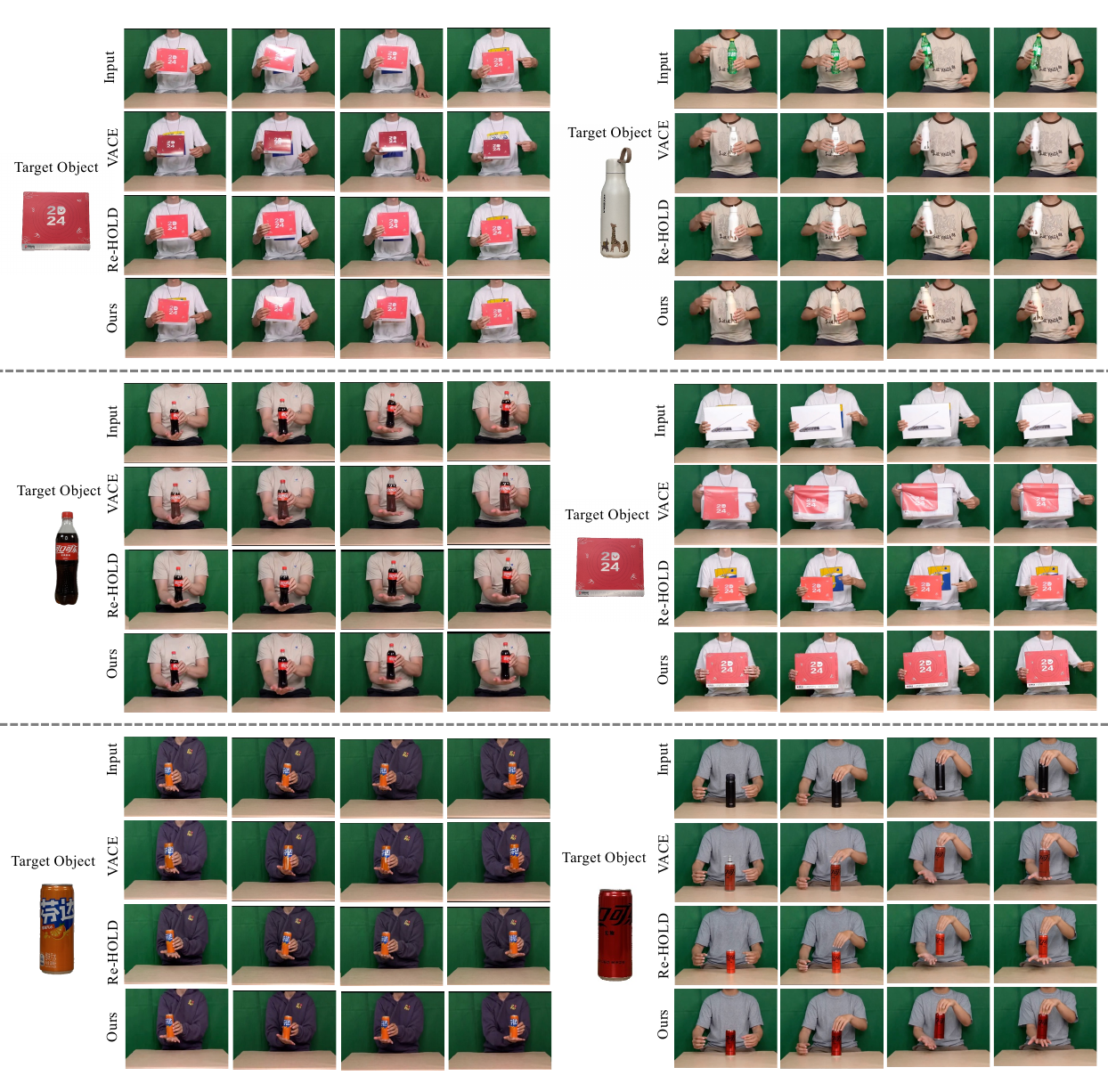}
\caption{
Additional self-reenactment (left column) and cross-reenactment (right column) videos generated by our method and baseline methods on the Re-HOLD dataset.}
\label{fig:exp1}
\end{figure*}

\begin{figure*}[t]
\includegraphics[width=\linewidth]{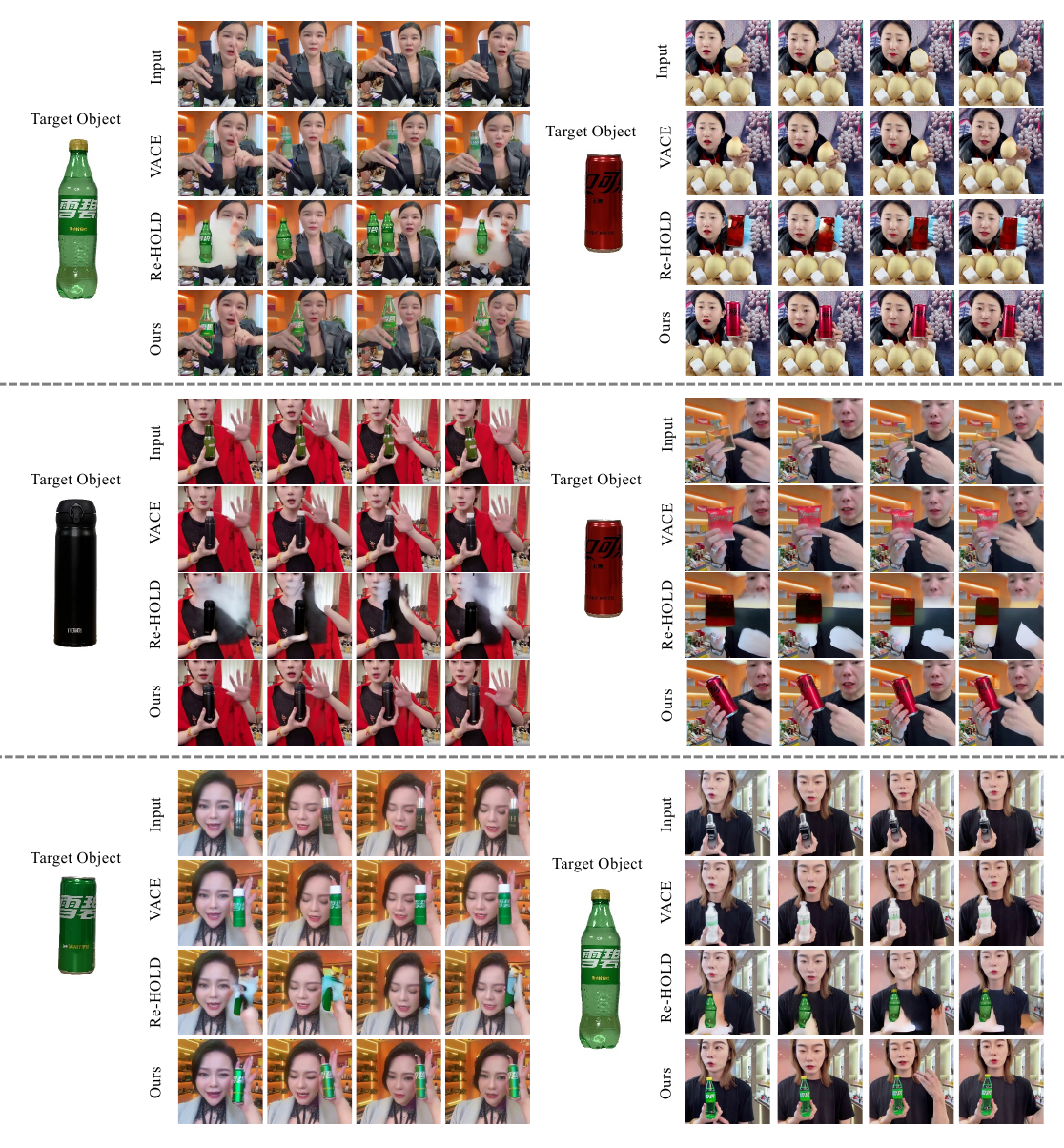}
\caption{
Additional cross-reenactment videos generated by our method and baseline methods on the HOI-ITW dataset.}
\label{fig:exp1}
\end{figure*}
\section{Experimental Results}

\begin{figure}[!t]
 \centering
 \includegraphics[width=0.97\linewidth]{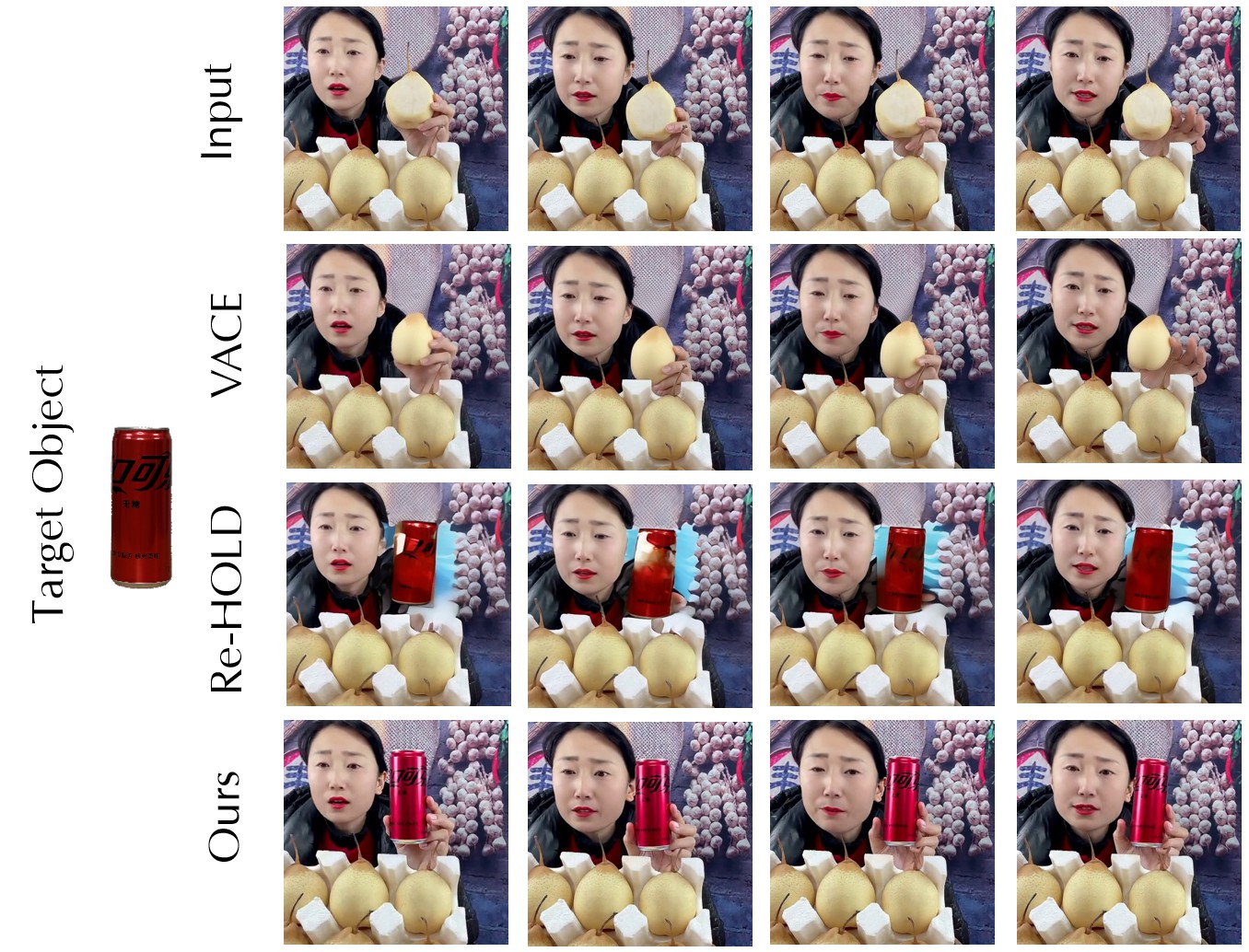} 
 \caption{\footnotesize Cross-reenactment results on the HOI-ITW dataset produced by our proposed method and other baseline methods.}
 \label{fig: wild}
\end{figure}
\subsection{Implementation Details}
\noindent \textbf{Implementation.} We implement the proposed DiT video generation model based on the pretrained Wan-14B~\cite{wan2025wanopenadvancedlargescale}-I2V model and FLUX.1-dev FLUX \cite{flux2024}. The models are trained in a full fine-tuning manner using their original loss functions. The training uses a dataset consisting of $19,000$ video clips on a server with 8$\times$ NVIDIA H100 80G GPUs. 

\vspace{0.5em}
\noindent \textbf{Dataset.} We evaluate the performance of the proposed method on two test datasets:
(1) Re-HOLD dataset: This dataset contains 139 videos for self-reenactment and 140 videos for cross-reenactment. It is used to assess the performance of our method on in-domain data.
(2) HOI-ITW (in-the-wild) dataset: This dataset consists of 30 videos for self-reenactment and 10 videos for cross-reenactment. All videos were collected in an e-commerce livestreaming environment, and we ensure that the data or similar data has not been seen by the trained models. This dataset is used to evaluate the generalization ability of the proposed method.

\vspace{0.5em}
\noindent \textbf{Baseline.} We conduct a comparative analysis of our method against a range of state-of-the-art (SOTA) techniques, including those specifically designed for HOI reenactment, such as Re-HOLD~\cite{fan2025re} and HOI-Swap~\cite{hoi-swap}, as well as general reference-guided video generation methods such as VACE~\cite{jiang2025vace}, AnimateAnyone~\cite{animateanyone}, AnyV2V~\cite{anyv2v}, RealisDance~\cite{realisdance}, and VideoSwap~\cite{videoswap}. This comprehensive evaluation allows us to assess the effectiveness of our proposed method from multiple perspectives.

\subsection{Evaluation Setting}
We evaluate the proposed method from two perspectives: (1) \textbf{Self-reenactment.} The method reconstructs the original video using the masked source video and the object image from the source. Following the Re-HOLD framework, we use PSNR and FID to evaluate the similarity between the reconstructed video and the original. Furthermore, we assess overall video quality through metrics of subject consistency and motion smoothness; (2) \textbf{Cross-reenactment.} The method substitutes the object in the original video with one not present in the source video, utilizing the masked source and a new object image. Continuing with the Re-HOLD framework, we assess overall video quality through metrics of subject consistency and motion smoothness.
However, the lack of ground-truth videos in this scenario complicates objective evaluation, prompting us to conduct a user study for quantitative comparison.

\vspace{0.5em}
\noindent \textbf{User Study.} We conducted a user study with 10 participants who evaluated 12 videos from the Re-HOLD dataset and 10 videos from the  HOI-ITW(in-the-wild) dataset. The participants rated the generated videos based on three criteria: (1) object consistency with the reference image, (2) motion consistency, and (3) overall video quality (independent of object consistency). Each criterion was assessed on a scale from 1 to 5, where 1 represents the lowest quality and 5 the highest. The user study was approved by the IRB.

\begin{figure}[!t]
 \centering
 \includegraphics[width=0.9\linewidth]{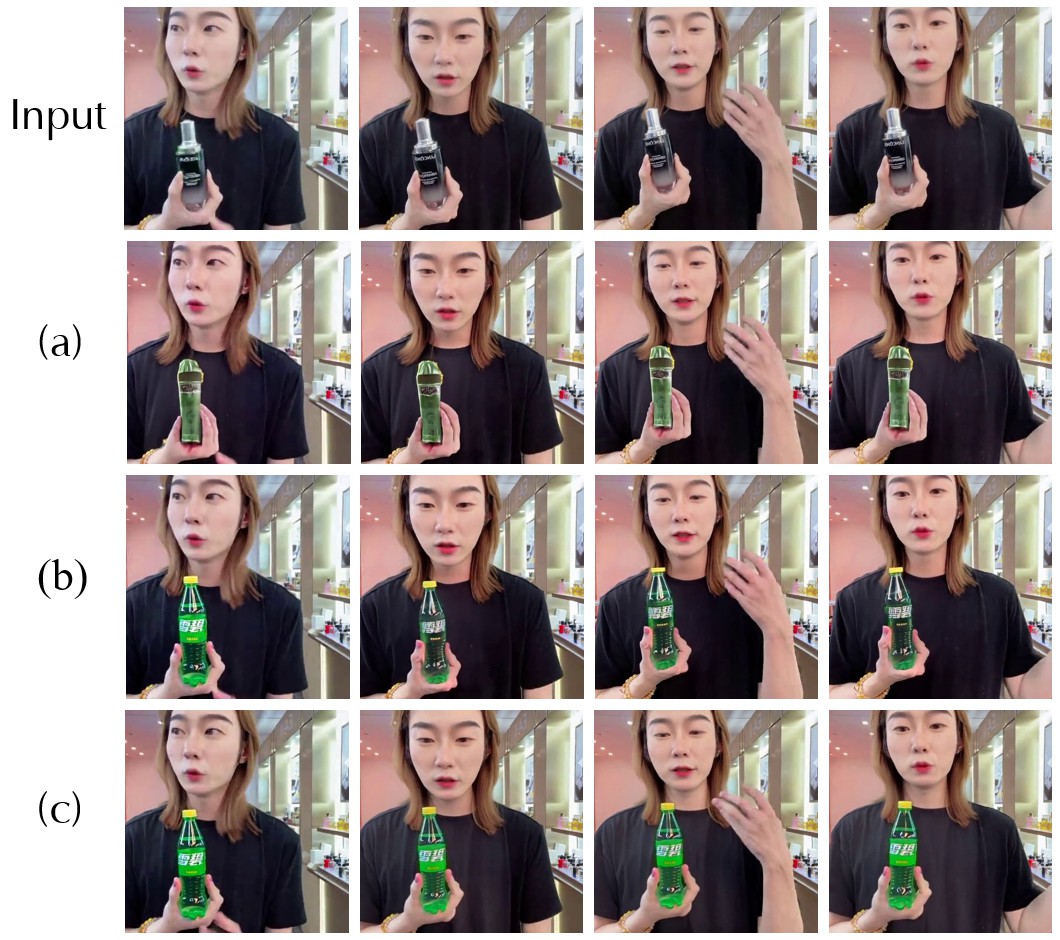} 
 \caption{\footnotesize Visualization results of the ablation study on the HOI-ITW dataset: (a) denotes no key frame generation; (b) denotes no object information fusion in Inp-TPU; (c) denotes results produced by our full model.}
 \label{fig:ablation}
\end{figure}
\subsection{Comparison with other methods}
\subsubsection{Evaluation on the Re-HOLD Dataset}
We evaluate the proposed method on the Re-HOLD dataset, covering cross-reenactment, self-reenactment, a user study, and qualitative comparisons. As shown in Tab.~\ref{tab:main_result}, our method achieves top performance in 4 out of 6 metrics and ranks second in 1 metric across both cross- and self-reenactment tasks. Tab.~\ref{tab:user_study} presents the user study results, where our method consistently outperforms competitors across all three evaluated dimensions: video quality, object consistency, and temporal consistency. These results validate not only the numerical superiority of our model but also its perceptual advantages as recognized by human evaluators.

Compared to the concurrent work of VACE, our approach exhibits a clear advantage in preserving reference fidelity—a critical aspect in object-centric video reenactment. As illustrated in the third row of Fig.~\ref{fig:exp1}, VACE fails to accurately preserve object attributes, mistakenly generating a pump head on the inferred Coke can. In contrast, our method preserves reference fidelity while simultaneously achieving high video quality and strong temporal consistency, demonstrating a better balance between fidelity and realism.

\begin{table*}[ht]
  \centering
  \caption{Evaluation on human-centric Re-HOLD and HOI-ITW datasets. * denotes that metrics are computed on a fixed length of 81 frames for fair comparison.}
  \label{tab:Re-HOLD-results}
  \fontsize{7.5pt}{9pt}\selectfont{%
  \begin{tabular}{llcc|cccc}
    \toprule
    \multirow{2}{*}{Dataset} & \multirow{2}{*}{Method}
      & \multicolumn{2}{c|}{Cross-Reenactment}
      & \multicolumn{4}{c}{Self-Reenactment} \\
    \cmidrule(lr){3-4} \cmidrule(lr){5-8}
     & 
      & subj. cons.\,$\uparrow$
      & mot. smth.\,$\uparrow$
      & PSNR\,$\uparrow$
      & FID\,$\downarrow$
      & subj. cons.\,$\uparrow$
      & mot. smth.\,$\uparrow$ \\
    \midrule
    \multirow{8}{*}{Re-HOLD}
      & AnyV2V         & 0.829 & 0.983 & 30.166 & 116.084 & 0.931 & 0.992 \\
      & VideoSwap      & 0.922 & 0.992 & 32.903 & 100.840 & 0.943 & 0.993 \\
      & AnimateAnyone  & \textbf{\B{0.950}} & 0.991 & 32.611 &  26.361 & 0.951 & 0.992 \\
      & RealisDance    & 0.948 & 0.991 & 32.784 &  26.337 & 0.951 & 0.993 \\
      & HOI-Swap       & 0.944 & \textbf{\B{0.994}} & 31.634 &  30.932 & 0.949 & 0.994 \\
      & (Re-HOLD)        & 0.955 & 0.994 & 33.451 &  19.021 & 0.953 & 0.995 \\
      & Re-HOLD\textsuperscript{*}     & \textbf{0.958} & \textbf{0.995} & 32.96 &  \textbf{\B{13.79}} & \textbf{0.958} & \textbf{\B{0.995}} \\
      & VACE\textsuperscript{*}           & 0.948 & \textbf{0.995} & \textbf{35.86} &  27.34 & \textbf{\B{0.952}} & \textbf{\B{0.995}} \\
      & Ours\textsuperscript{*}           & 0.948 & \textbf{0.995} & \textbf{\B{33.74}} &  \textbf{12.07} & \textbf{0.958} & \textbf{0.996} \\
    \midrule
    \multirow{3}{*}{HOI-ITW}
      & VACE\textsuperscript{*}           & \textbf{0.954} & \textbf{0.989} & \textbf{33.12} &  \textbf{\B{59.94}} & \textbf{0.946} & \textbf{\B{0.988}} \\
      & Re-HOLD\textsuperscript{*}        & 0.923 & 0.987 & 30.35 &  127.8 & 0.901 & 0.987 \\
      & Ours\textsuperscript{*}\footnotemark           & \textbf{\B{0.953}}   & \textbf{0.989}   & \textbf{\B{31.48}} &  \textbf{35.01} & \textbf{0.946} & \textbf{0.989} \\
    \bottomrule
  \end{tabular}%
  }
  \label{tab:main_result}
\end{table*}
\subsubsection{Evaluation on the HOI-ITW Dataset}

We further evaluate the proposed method on the HOI-ITW dataset. The experimental settings are more challenging and closely aligned with practical applications. As shown in Tab. \ref{tab:main_result}, the performance gap between our method and existing baselines becomes even more pronounced in unseen scenes. Our method consistently demonstrates high reference fidelity and temporal consistency, while competing approaches often suffer from degraded visual quality and reduced reference alignment. This highlights the superior robustness and adaptability of our method to novel domains—an essential quality for real-world deployment.

Notably, the previous state-of-the-art method, Re-HOLD, completely fails in the unseen setting, achieving only 1.1, 1.7, and 1.1 in 3 aspects of the user study. Another baseline, VACE, performs reasonably well in video quality and temporal consistency, but its reference fidelity drops significantly—from 2.78 on the Re-HOLD dataset to 1.55 in the unseen scenario. This degradation is primarily due to VACE’s limited ability to disentangle the background and reference image inputs. As illustrated in Fig. \ref{fig: wild}, VACE generates an object (e.g., a pear) based on background cues, entirely ignoring the given target object. In contrast, our method successfully disentangles background and reference input, enabling faithful adherence to the target object during video generation.

Overall, our method achieves top performance in 7 out of 9 evaluation metrics and ranks second in the remaining two, as shown in Tab. \ref{tab:main_result} and \ref{user_study}. The combination of strong quantitative results, positive user feedback, and high visual fidelity positions our approach as the new state-of-the-art in reference-based human-object video reenactment.

\begin{table}[!ht]
  \centering
  \caption{User study of various methods across different datasets.}
  \label{tab:user_study}
  \fontsize{7.5pt}{9pt}\selectfont{%
    \begin{tabular}{llccc}
      \toprule
      Dataset & Method           & Video quality & Ref. Fidelity & Temporal cons. \\
      \midrule
      \multirow{3}{*}{Re-HOLD}
      & VACE & 3.65   & 2.78      & 3.72        \\
      & Re-HOLD & 3.40   & 3.73      & 3.30        \\
      & Ours & \textbf{3.80}   & \textbf{3.87}      & \textbf{3.90}        \\
      \midrule
      \multirow{3}{*}{HOI-ITW}
      & VACE & 3.275   & 1.55      & 3.575        \\
      & Re-HOLD & 1.125   & 1.7      & 1.1        \\
      & Ours & \textbf{3.525}   & \textbf{3.875}      & \textbf{3.55}        \\
      \bottomrule
    \end{tabular}%
  }
  \label{user_study}
\end{table}

\subsection{Ablation Study}

To verify the effectiveness of different modules, we set a series of experiments in this section. We use the HOI-ITW dataset for testing and two types of experiments have been executed here:
\vspace{0.05in}

\noindent \textbf{No Key Frame Generation.} We evaluate the impact of removing the key frame generation stage by directly generating the final result using a one-stage approach. As shown in Tab.~\ref{tab:ablation}, all metrics degrade without the initial key frame generation. Visualizations in Fig.~\ref{fig:ablation} (a) further support this observation. As illustrated, the model struggles to follow the target object in the absence of key frame guidance.

\vspace{0.05in}
\noindent \textbf{No Object Information Fusion in Inp-TPU.} We also examine the effect of removing object information fusion in the Inp-TPU module. In this setting, subsequent frame generation relies solely on the key frame for object guidance. As reported in Tab. \ref{tab:ablation}, this results in a 27.08\% drop in FID. A similar performance degradation is observed in cross-reenactment tasks. Fig.~\ref{fig:ablation} (b) provides qualitative evidence, showing that the target object undergoes more noticeable changes across frames when object information is not fused.

\begin{table}[!t]
  \centering
  \caption{Ablation study on the impact of generating key frames and implementing our proposed Inp-TPU on the HOI-ITW dataset.}
  \label{tab:ablation}
  \fontsize{7.5pt}{9pt}\selectfont{%
    \begin{tabular}{lcccc}
      \toprule
      \multirow{2}{*}{Variations}
        & \multicolumn{2}{c}{Cross-Reenactment}
        & \multicolumn{2}{c}{Self-Reenactment} \\
      \cmidrule(lr){2-3} \cmidrule(lr){4-5}
        & subj.\ cons.\,$\uparrow$
        & mot. smth.\,$\uparrow$
        & PSNR $\uparrow$
        & FID $\downarrow$ \\
      \midrule
      - First frame
        & 0.94 & 0.990 & 30.24 & 81.81 \\
      - Ref. img.
        & 0.92    & 0.990    & 30.53 & 48.25 \\
      \midrule
      \textbf{Full}
        & \textbf{0.95} & \textbf{0.990}
        & \textbf{31.48} & \textbf{35.01} \\
      \bottomrule
    \end{tabular}%
  }
\end{table}

\section{Conclusion}

In conclusion, this paper introduces a novel framework for high-fidelity Hand-Object Interaction (HOI) video generation, addressing key challenges such as occlusion and variations in object dynamics. By utilizing a unified inpainting-based token process and a two-stage video diffusion transformer, our method effectively generates realistic HOI reenactments, including in-the-wild scenarios. The ability to leverage pretrained context perception without adding parameters enhances generalization to unseen objects and humans. Comprehensive evaluations confirm that our approach outperforms existing methods, delivering superior realism and seamless interactions.


\noindent \textbf{Limitation.} While our method achieves state-of-the-art performance among existing approaches, it has the following limitations: 1) Currently, it operates as a two-stage pipeline, requiring the model to denoise twice, unlike a single-stage pipeline. We aim to explore this in the future; 2) Our method generates the reenactment video based solely on the object image and masked video, limiting fine-grained manipulations such as object rotation and flipping. This limitation could be addressed by incorporating 6D object pose information~\cite{wen2024foundationpose} in future work.

\section*{Acknowlodgement}
The authors would like to thank Xuan Huang, Xing Liu, Chunyu Song, Jinbo Wu, Xiaobo Gao, Quanwei Yang, Yingying Fan, Ruoyu Wang, Yu Wang, and Jiazhi Guan for their support in preparing the experiments.
{
    \small
    \bibliographystyle{ieeenat_fullname}
    \bibliography{main}
}

\end{document}